\documentclass[12pt]{iopart}
\usepackage{graphicx}
\usepackage{subfigure}

\begin{document}

\title{Strange particle production in a single-freeze-out model}

\author{W.~Florkowski$^{1,2}$, W.~Broniowski$^1$, and A.~Baran$^1$}

\address{$^1$ The H. Niewodnicza\'nski Institute of Nuclear Physics, 
Polish Academy of Sciences, 31-342 Krak\'ow, Poland} 
\address{$^2$ Institute of Physics, \'Swi\c{e}tokrzyska Academy,
25-406 Kielce, Poland}
 
\begin{abstract}
The transverse-momentum spectra and elliptic flow of strange particles
are calculated in the framework of the single-freeze-out model and compared
with the recent RHIC data. An overall very good agreement between the 
model predictions and the data is obtained.
\end{abstract}

In this paper we  use the single-freeze-out model of Refs. 
\cite{Broniowski:2001we,Broniowski:2001uk,Broniowski:2002nf} to analyze the
production of strange particles in relativistic heavy-ion collisions. The model
results are compared with the recent RHIC data on the transverse-momentum spectra
and elliptic flow of hyperons \cite{Adams:2003fy,Castillo:2004jy}.
In the original formulation \cite{Broniowski:2001we,Broniowski:2001uk,Broniowski:2002nf}, 
the single-freeze-out model is used to describe cylindrically symmetric sytems such as 
those created in the most central events.  
A natural extension of the original formulation is obtained by the introduction of the 
azimuthal asymmetry in both the shape of the freeze-out hyper-surface and the transverse-flow 
field. The asymmetry of the freeze-out hypersurface is taken into account by the following 
parameterization \cite{Broniowski:2002wp}:
\begin{eqnarray}
t &=& \tau\, \hbox{cosh} \alpha_\parallel \, \hbox{cosh} \alpha_\perp , \quad 
r_z = \tau\, \hbox{sinh} \alpha_\parallel \, \hbox{cosh} \alpha_\perp 
, \nonumber \\
r_x&=& \tau\, \hbox{sinh} \alpha_\perp  \sqrt{1-\epsilon} \,\cos \phi , \quad
r_y = \tau\, \hbox{sinh} \alpha_\perp   \sqrt{1+\epsilon} \,\sin \phi.
\label{rmod}
\end{eqnarray}
Here the coordinates $t$ and $r_z$ are kept as in the symmetric formulation:
$\tau$ denotes the lifetime of the system, whereas $\alpha_\parallel$ and
$\alpha_\perp$ are the longitudinal and transverse space-time rapidity.
The $r_x$-axis lies in the reaction plane, and the $r_y$-axis
is perpendicular to the reaction plane. The almond-like shape of the system in
the transverse plane is achieved by the introduction of the factors 
$\sqrt{1 \pm \epsilon}$. For positive $\epsilon$ the system is
elongated out of the reaction plane, as seen in the experiment. 
The value of $\epsilon$ is fitted to reproduce the dependence of the $R_{\rm side}$
HBT radius on the azimuthal angle \cite{Broniowski:2002wp}.
The four-velocity field is defined in the following way:
\begin{eqnarray}
u_t = \frac{t}{N},  \quad u_z = \frac{r_z}{N}, \quad
u_x = \frac{r_x}{N} \sqrt{1+\delta} \,\cos \phi, \quad 
u_y =  \frac{r_y}{N} \sqrt{1-\delta} \,\sin \phi.
\label{umod}
\end{eqnarray}
Here the normalization $N$ is taken in such a way that the four-velocity field
is properly normalized, $u^\mu u_\mu=1$. Positive $\delta$ means a faster flow in 
the reaction plane, which corresponds to positive $v_2$. Certainly, the choice (\ref{rmod},
\ref{umod}) is by no means unique, but it takes into account the essential 
features found in the experiments.

\begin{figure}[b]
\begin{center}
\includegraphics[width=9cm]{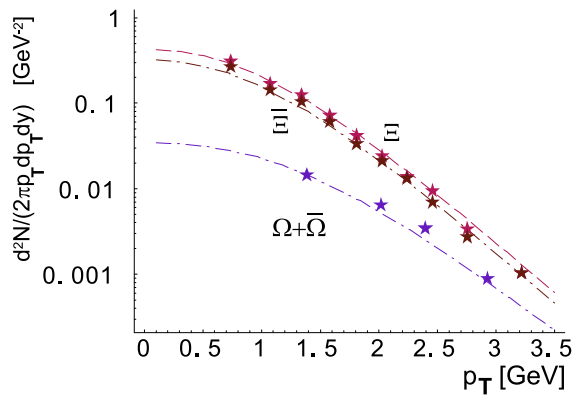}
\end{center}
\caption{Comparison of our prediction \cite{Broniowski:2001uk}
on the production of $\Xi$ and $\Omega+{\bar \Omega}$
in Au-Au collisions at the energy $\sqrt{s_{NN}} = $ 130 GeV 
with the final data for most central collisions
published by the STAR Collaboration \cite{Adams:2003fy}. }
\end{figure}

The transverse-momentum spectra are calculated from the Cooper-Frye formula
\cite{Cooper:1974mv}
\begin{equation}
\frac{dN}{d^{2}p_{\perp }dy} 
=
\int p^{\mu }d\Sigma _{\mu }\ f\left(p\cdot u\right),
\label{Ni}
\end{equation}
where $d\Sigma_\mu$ denotes the element of the hypersurface defined by Eq. (\ref{rmod}),
the four-velocity $u$ is defined by Eq. (\ref{umod}), and the hadron 
distribution functions $f$ include sequential decays of the resonances, as
described in a greater detail in Refs. \cite{Broniowski:2001uk,Broniowski:2002nf}.
Then, the coefficient of the elliptic flow is obtained from the formula
\begin{equation}
v_2(p_\perp) = { \int\limits_0^{2\pi} d\phi \frac{dN}{d^{2}p_{\perp }dy} \cos(2\phi)
\over  \int\limits_0^{2\pi} d\phi \frac{dN}{d^{2}p_{\perp }dy} }.
\label{v2}
\end{equation}

We note that the 
parameters of the model may be classified into three independently-fitted groups: the two 
{\it thermodynamic parameters} (temperature $T$ and the baryon chemical potential $\mu_B$) 
are obtained from the study of the ratios of hadronic abundances \cite{Florkowski:2001fp}; 
the two {\it geometric parameters } (the lifetime of the system $\tau$ and the radius of 
the firecylinder $\rho_{\rm max} = \tau \, \hbox{sinh} \alpha^{\rm max}_\perp$)  
are obtained from the study of the azimuthally averaged transverse-momentum spectra; 
and, finally, the two {\it deformation parameters} ($\epsilon$ and $\delta$) 
are obtained from the experimental information on $R_{\rm side}$ and the $v_2$ 
coefficient.

It is important to stress that in the fits of the geometric and deformation parameters we 
use only the data for the most abundant particles (pions, kaons, protons and antiprotons). 
In this way, the calculation of the 
transverse-momentum spectra or $v_2$ of strange particles may be regarded as the prediction 
of the model. An example of such a procedure is the calculation of the 
transverse-momentum spectra of hyperons \cite{Broniowski:2001uk}. In Fig. 1 we can see the 
comparison of the predicted spectra for the energy $\sqrt{s_{NN}} = $ 130 GeV
with the final results published  recently by the STAR 
Collaboration  \cite{Adams:2003fy}. In this case the most central collisions are taken into 
account, hence $\epsilon=0$ and $\delta=0$. The fit of the remaining parameters gives: 
$T =$ 165~MeV, $\mu_B=$ 41~MeV, $\tau = 7.66$ fm, and $\rho_{\rm max}$ = 6.69 fm 
\cite{Broniowski:2001uk}. We can observe a very good agreement of the model predictions
with the data. A similar comparison for other particles can be found in 
\cite{Broniowski:2001uk,Broniowski:2002wp2}.
The overall agreement is satisfactory although some discrapancies for the case of
Lambdas (normalization) or antiprotons may be noticed.
\begin{figure}[b]
\begin{center}
\includegraphics[width=10.5cm]{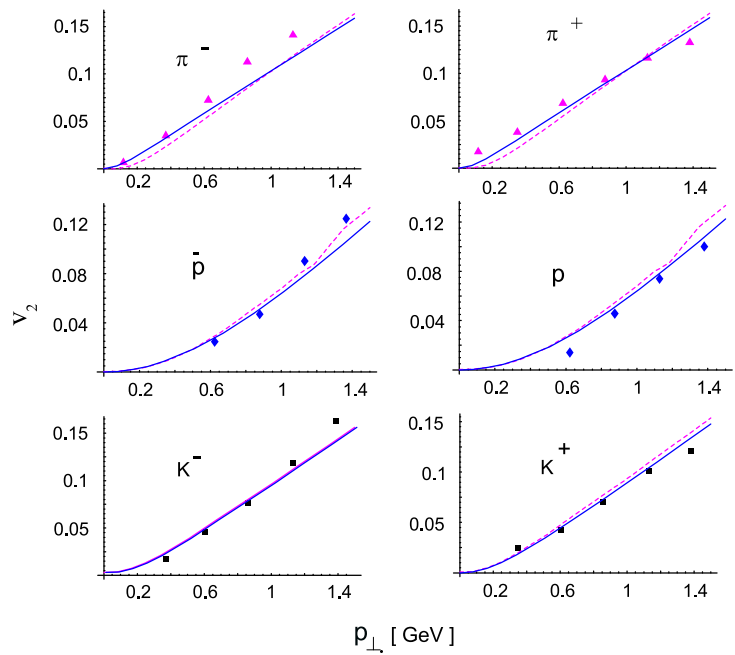}
\end{center}
\caption{The fit of the elliptic-flow coefficient $v_2$ for pions, kaons,
and (anti)protons. The symbols represent preliminary minimum bias data from the
PHENIX Collaboration collected at the energy $\sqrt{s_{NN}}$ = 200 GeV (presented
by Voloshin in Ref. \cite{Voloshin:2002wa} ) .}
\end{figure}

In Fig. 2 we show the fit to the PHENIX minimum-bias data for the energy 
$\sqrt{s_{NN}} = $ 200 GeV showing the $p_\perp$-dependence
of the $v_2$ coefficient of pions, kaons, and (anti)protons \cite{Voloshin:2002wa}.
In this case the values of the thermodynamic parameters  are: 
$\mbox{$T$ = 165 MeV}$ and $\mu_B = 26$ MeV, 
whereas the values of the geometric parameters are $\tau = 4.04$ fm and $\rho_{\rm max} = 
3.90$ fm \cite{Baran:2003nm,AnnaBaranPhD}. The new experimental information 
contained in the $p_\perp$-dependence of the $v_2$ coefficient allows us to fit 
the value of $\delta$. In the considered case we find 
$\delta = 0.25$, while $\epsilon = 0.13$ has been read off from the 
azimuthally-dependent HBT data \cite{Broniowski:2002wp}.
The solid lines in Fig. 2 correspond to the calculations
done without the inclusion of the sequential hadronic decays, while the dotted lines
describe the results obtained with the full feeding from such decays. We conclude
that the effects of the resonance decays are not important for the good 
description of the elliptic-flow coefficient $v_2$ (nota bene, the effects of 
the hadronic decays are crucial for good description of the $p_\perp$-spectra).
A very good descrition of the discussed data has been achieved by Hirano and Tsuda 
in their full hydrodynamical calculation \cite{Hirano}.

The values of the parameters obtained in the analysis of the production of 
the most abundant particles may be used to obtain the function $v_2(p_\perp)$ for 
strange particles. In Fig. 3
we show a comparison of the model calculation with the data representing
the elliptic flow of the Lambda ($\Lambda + {\bar \Lambda}$) and, what is
especially interesting, of the multistrange baryons \cite{Castillo:2004jy}.
In both cases we find a very good agreement. 

\begin{figure}[t]
\begin{center}
\subfigure{\includegraphics[angle=0,width=0.49\textwidth]{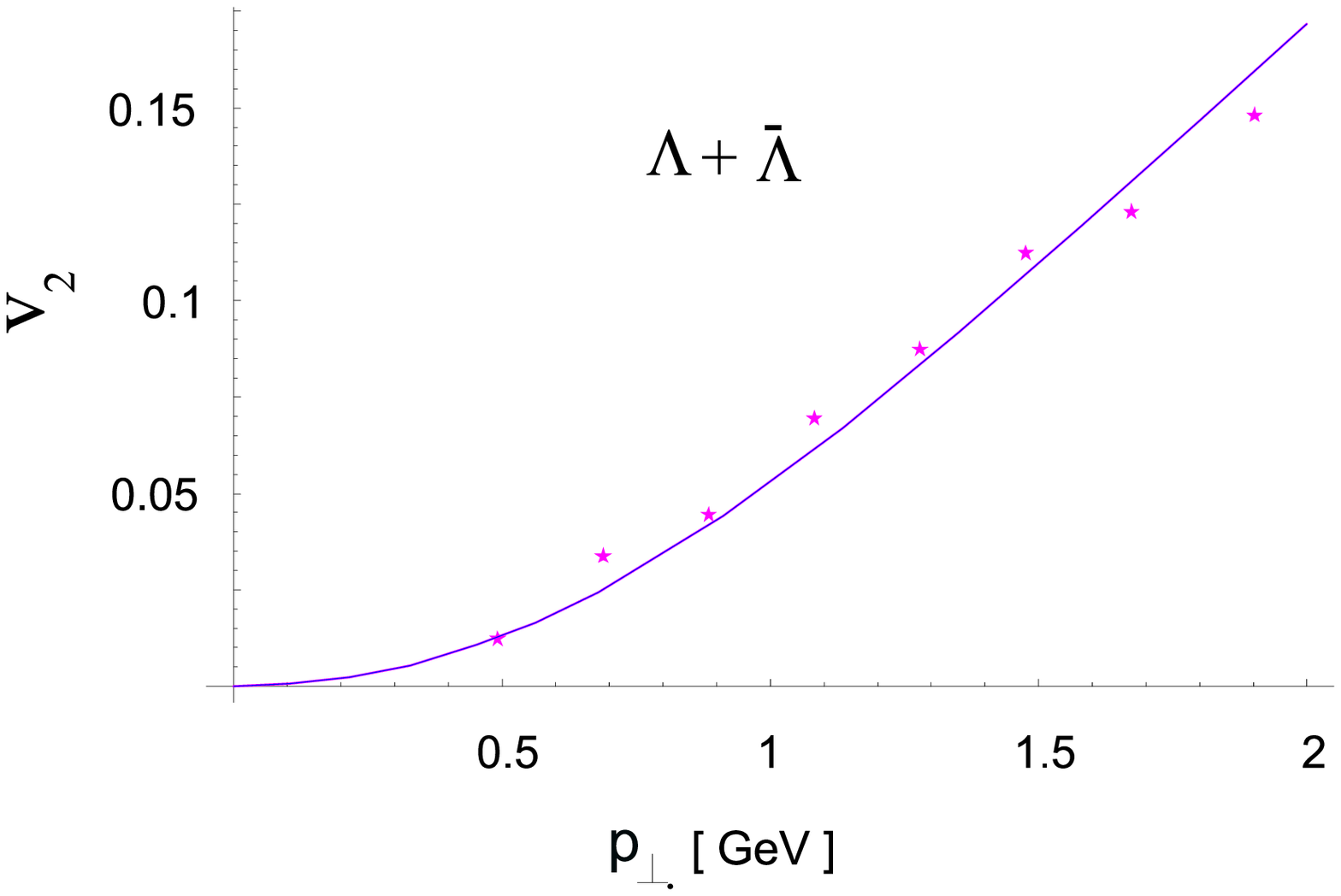}} 
\subfigure{\includegraphics[angle=0,width=0.49\textwidth]{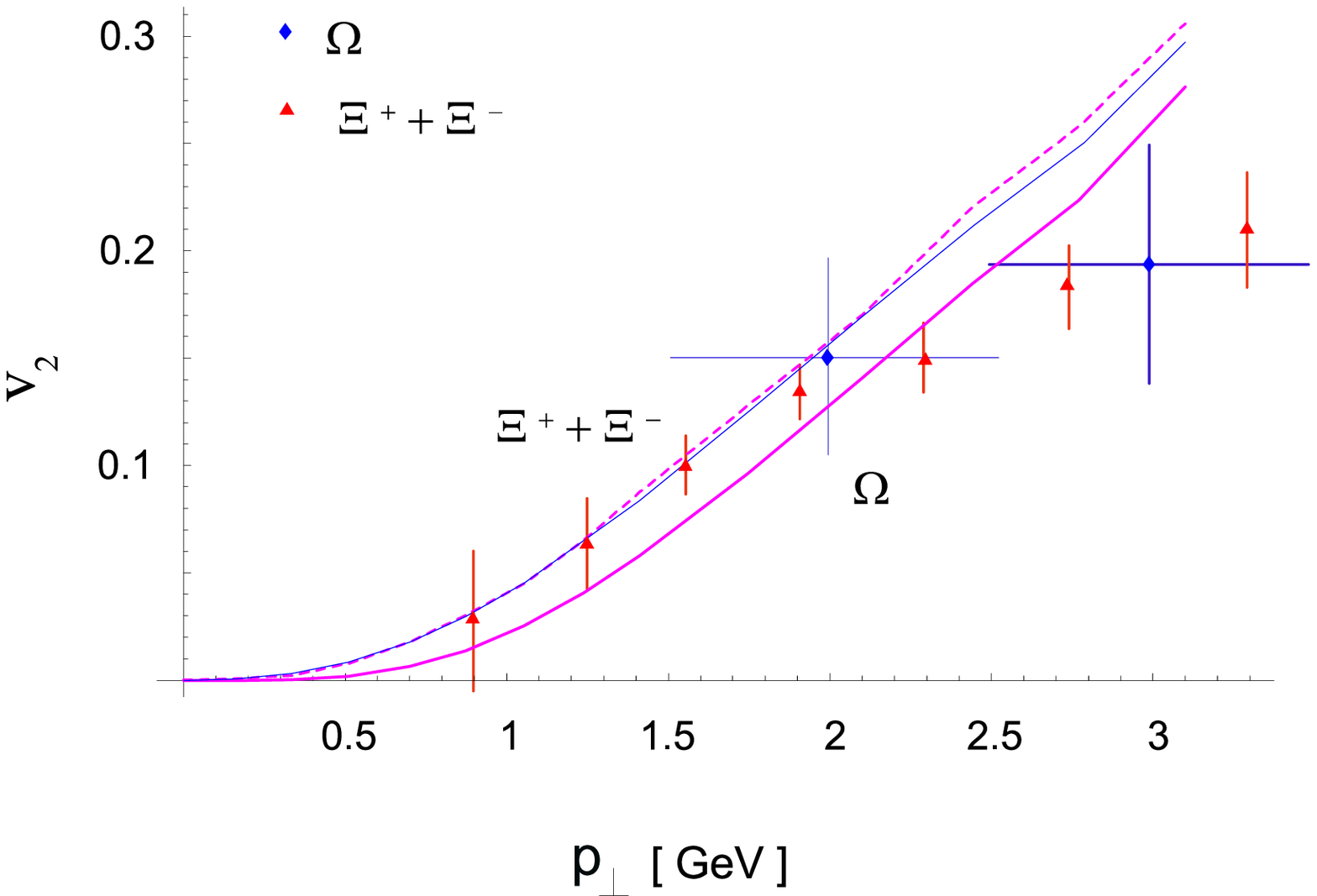}} \\
\end{center}
\caption{The $v_2$ coefficient for the sum $\Lambda+{\bar \Lambda}$ (left) and 
multistrange baryons (right). The points show the minimum-bias data from STAR 
collected at the beam energy $\sqrt{s_{NN}}$ = 200 GeV. The model parameters are
the same as those obtained in the analysis of the most abundant 
particles measured in the 
minimum-bias events by PHENIX, shown in Fig. 2.}
\end{figure}

As a conclusion, we would like to emphasize that the single freeze-out model
describes in the unified way the production of ordinary and strange hadrons
in the soft region $p_\perp < 2$ GeV.
The values of the parameters extracted from the pion, kaon, and (anti)proton
data  allow us
to make predictions about the behavior of strange particles. Such
predictions give a satisfactory description of the experimental data.

This research was supported in part by the Polish State Committee for
Scientific Research, grant number 2~P03B~05925.


\end{document}